\documentclass[11pt]{article}
\usepackage{graphicx}
\begin{document}

\title{Revisiting the OZI-forbidden Radiative
Decays of Orthoquarkonia\footnote{Supported by National Natural
Science Foundation of China.} } \vspace{3mm}

\author{{Gang Li$^1$, Tong Li$^2$, Xue-Qian Li$^2$, Wen-Gan Ma$^1$ and Shu-Min Zhao$^2$}\\
{\small $^{1}$ Modern Physics Department, University of Science and Technology of China} \\
{\small  (USTC), Hefei, Anhui 230026 China}\\
{\small $^{2}$ Department of Physics, Nankai University, Tianjin
300071, China} }
\date{}
\maketitle \vskip 12mm

\begin{abstract}
It is interesting to investigate the OZI-forbidden radiative
decays of orthoquarkonium, $J/\psi\rightarrow
\gamma\pi^0,\;\gamma\eta$ and $\gamma\eta'$ in perturbative QCD.
In this work without the approximations adopted in literature we
carry out a full one-loop calculation which involves integrations
of 4-point and 5-point loop functions. Our numerical results are
in agreement with the present data. We also briefly discuss the
decays of $J/\psi\rightarrow \gamma+\rho^0$, as well as
$\Upsilon\rightarrow \pi^0\gamma,\;\eta\gamma, \eta'\gamma$.
\end{abstract}

\section{Introduction}

The OZI rule\cite{OZI} plays an important role in the processes
which occur via strong interaction and in general the concerned
calculations are carried out in the framework of perturbative QCD.
Thus careful studies on such processes where the OZI rule applies
can deepen our insight to the perturbative QCD. On other aspect,
we would encounter another serious problem. Namely, even though we
can accurately calculate the processes at quark-gluon level in
terms of perturbative QCD, for evaluating the transition matrix
elements, the hadronization which is related to the
non-perturbative QCD, must be dealt with. Because of lack of solid
knowledge on the non-perturbative QCD, for a whole calculation,
one has to employ some model-dependent wavefunctions, which may
contaminate the theoretical results. Thus to reduce theoretical
uncertainties, the wavefunctions should be well tested and
theoretically studied. Then, although the results have more or
less model-dependence, one can trust that they are comparatively
reliable. Moreover, as widely discussed, the radiative decays of
$J/\psi$ may provide an ideal place to look for the mysterious
glueballs \cite{Close}. Since the structure of glueballs is not
clear so far, we ignore contributions from possible glueball
resonances and just concentrate our focus on the direct processes
which are supposed to correspond to the experimental data.

K\"{o}rner et al.\cite{Korner} investigated the OZI-forbidden
radiative decays of orthoquarkonia in perturbative QCD and their
pioneer work was done more than 20 years ago. Since then, the
technique of calculating loop diagrams has been improved and the
knowledge on the wavefunctions of light mesons such as $\pi$,
$\eta,\;\eta'$ is much enriched. Meanwhile the corresponding
experimental measurements become more precise and make it possible
to test our theoretical understanding on both the perturbative QCD
calculation and the hadron-wavefunctions which are overwhelmingly
governed by non-perturbative QCD.

In K\"{o}rner et al.'s work \cite{Korner}, the authors tactfully
dealt with the complicated Feynman integrations of four-point and
five-point loop functions, i.e. the D- and
E-functions\cite{loop,Dfunction}. They took the weak-binding
approximation for both heavy and light mesons, indeed generally
this approximation is reasonable and much simplifies the
calculations. By the approximation, the heavy quarks $Q$, $\bar Q$
($c$ and $\bar c$ for $J/\psi$) and light quarks $q$, $\bar q$ in
the decay product$-$light orthoquarkoinum are set to possess equal
momenta and be on their mass shells, i.e. $p_{Q}=p_{\bar Q}$,
$p_{q}=p_{\bar q}$ and $p_{Q}^2=m_{Q}^2$, $p_{q}^2=m_q^2$.

In the weak-binding approximation where $q$ and $\bar q$ have the
same momentum and are on mass shell, the flavor dependence of the
propagator disappears due to the on-shell condition, thus the
difference of flavors does not manifest in the integration. Since
$\pi^0$ is of structure ${1\over \sqrt 2}(d\bar d-u\bar u)$,
$\eta_8$ is of structure ${1\over\sqrt 6}(u\bar u+d\bar d-2s\bar
s)$, the contributions from different flavors cancel each other
and make the total width null, whereas only for
$\eta_0={1\over\sqrt 3}(u\bar u+d\bar d+s\bar s)$, all
contributions are added up and the result is non-zero. Therefore
to get non-zero widths for $J/\psi\rightarrow
\gamma\pi^0,\;\gamma\eta_8$, one must abandon the weak-binding
approximation, namely, consider the relative momentum between $q$
and $\bar q$ and not let them be on mass shell.

Noting the disadvantage of this approximation, Yang
\cite{Yang,CHuang} dismissed it and re-derived relevant formulas.
In his work where he neglected the mass of the light quark mass of
the light meson and obtained an improved analytical expression for
the rates of $J/\psi\rightarrow \eta\gamma,\;\eta'\gamma$.

As $J/\psi$ is much heavier than the produced light mesons
($\pi^0,\;\eta,\;\eta'$), the 3-momentum of the mesons is large,
namely much larger than the relative momentum between $q$ and
$\bar q$, the equal-momentum approximation as well as the
approximate on-shell condition are reasonable. However, this
approximation probably too simplifies the picture. Concretely,
$J/\psi$ (or $\Upsilon$) is an SU(3) singlet, $\pi^0$ belongs to
isospin-1, $\eta$ has a large fraction of $\eta_8$, and in
$\eta'$, the $\eta_0$ component dominates. Strictly, the
electromagnetic interaction does not demand an isospin
conservation, i.e. the isospin of photon can be either 0 or 1. But
as we take the most probable structure that isospin of photon is
0, the $J/\psi\rightarrow \gamma\pi^0$ is an isospin-violating
process and $J/\psi\rightarrow \gamma\eta_8$ is an SU(3) violating
process and only $J/\psi\rightarrow \gamma\eta_0$ is an SU(3)
conserving process. One can expect the sequence $\Gamma(
J/\psi\rightarrow \gamma\pi^0)<\Gamma(J/\psi\rightarrow
\gamma\eta_8)<\Gamma(J/\psi\rightarrow \gamma\eta_0)$ and the
present experimental data on $J/\psi$ confirm this conjecture
\cite{data0}. In our calculations, we will show that this sequence
is completely understood in a full loop calculation, i.e. the
cancellation among different quark-flavors in $\pi^0$ and $\eta_8$
causes this pattern.

We are motivated to re-evaluate the processes $J/\psi\rightarrow
\gamma\pi^0,\;\gamma\eta,\;\gamma\eta'$ by keeping an arbitrary
relative momentum between $q$ and $\bar q$ and non-degenerate
masses of the light quarks ($u,d,s$). We notice that just because
of this mass difference the three modes have different non-zero
branching ratios. In the early work \cite{Korner}, the authors
supposed that all processes occur via $J/\psi\rightarrow
\eta_0+\gamma$, and the mixing of $\eta_8$ and $\eta_0$ results in
the non-zero rate for $J/\psi\rightarrow\eta\gamma$.

However, as dismissing the weak-binding approximation, the
advantages for simplifying the calculations are lost, namely, one
cannot approximately reduce the five-point and four-point loop
functions into simple three-point loop functions. It is easy to
understand that under the weak-binding approximation where $q$ and
$\bar q$ have the same momentum and are on shell, the numerators
of the integrands can be properly decomposed into several groups
and each of them can cancel certain factors of the denominators,
so that the number of the Feynman parameters can be reduced,
whereas without the approximation, generally all the terms of the
5-point and 4-point loop functions remain and the expressions
cannot be further simplified (see below for details.).

Following the standard procedure \cite{loop}, we deal with the
5-point loop functions and then in term of the program
\cite{program}, we evaluate integrations of the 4-point and
3-point loop functions to obtain an effective vertex
\cite{coupling} for $J/\psi\rightarrow \gamma gg\rightarrow
\gamma+P$ where $P$ is a pseudoscalar.

Moreover, there seems to be another contribution from the tree
diagram
$$c\bar c(J/\psi)\rightarrow \gamma^*\rightarrow \gamma+q\bar
q(P)$$ shown in Fig.1 (a) and (b). By a simple analysis, one can
immediately note that the contributions from (a) and (b) exactly
cancel each other as long as the wavefunction of P is symmetric to
the two light-quark constituents. Therefore in the radiative
decays of orthoquarkonia, the leading contribution comes from the
one-loop OZI forbidden processes\cite{Korner,Yang}.

To obtain the decay amplitude, one  needs to evaluate the hadronic
matrix elements
\begin{equation}\label{hadron}
<P|V_{eff}|J/\psi>,
\end{equation}
where $P$ stands for $\pi^0,\;\eta$ and $\eta'$. In the
calculations, we use the light-cone wavefunction
\cite{wave1,wave2,light-cone2} for $P$.

In the non-relativistic model, the wavefunction of the light meson
$P$ at zero-point $R_{PS}(0)$ is responsible for the hadronization
effects\cite{Korner,coupling}.

The same problem has also been investigated by some other authors
\cite{Novikov,Ma}, especially an anomaly for $\Upsilon\rightarrow
\gamma\eta'$ is addressed. In their work, the gluon contents in
$\eta$ and $\eta'$ are considered and non-perturbative effects are
taken into account. In the work \cite{Ma}, the anomaly seems to be
alleviated in this approach, but the concerned assumption is still
under suspicion\cite{Yellow}. We will come back to this point
again in our last section.

The paper is organized as follows. After this long introduction,
in Sec.II, we present our formulation  and in Sec. III, we make
the numerical evaluation of the decay rates of $J/\psi\rightarrow
\gamma\pi^0,\gamma\eta,\gamma\eta'$ and some necessary input
parameters are explicitly given. To investigate the whole
scenario, we further calculate the rates of $J/\psi\rightarrow
\rho^0\gamma$ as well as $\Upsilon\rightarrow
\pi^0\gamma,\;\eta\gamma,\;\eta'\gamma$. We find an anomaly for
$\Upsilon\rightarrow \eta'\gamma$. The last section is devoted to
a simple discussion and our conclusion. Some of the tedious
details are collected in the appendices.

\section{Formulation of process $V \to P \gamma$}
\subsection{Derivation of the effective operators}

(i) For $V\rightarrow P\gamma$.

First we derive the formula for $J/\psi\rightarrow P\gamma$, where
$P$ stands as a neutral pseudoscalar meson ($\pi^0,\;\eta$ and
$\eta'$). As mentioned above, the tree diagrams shown in Fig.1 (a)
and (b) make null contributions due to a mutual cancellation among
them. The corresponding Feynman diagrams which are  OZI suppressed
and offer the leading contributions to the processes, are shown in
Fig.2 (a) through (f).

The effective vertex for $J/\psi\rightarrow \gamma+P$ where $P$
stands for a pseudoscalar is in the form \cite{Chung,effective
vertex}
\begin{equation}\label{eff}
V_{eff}=g_{eff}\epsilon_{\alpha\beta\mu\nu}p_{_V}^{\alpha}\varepsilon_{J/\psi}^{\beta}
q^{\mu}\varepsilon_{\gamma}^{*\nu},
\end{equation}
where $p_{_V}^{\alpha}=p_{_V}^0=M_{J/\psi}$ is the four-momentum
of $J/\psi$ in its center-of-mass frame,
$\varepsilon_{J/\psi}^{\beta}$ and $\varepsilon_{\gamma}^{*\nu}$
are the polarizations of $J/\psi$ and the emitted photon, $q$ is
the relative momentum of the photon and the pseudoscalar. The
$g_{eff}$ is an effective coupling and should be derived by
evaluating the corresponding Feynman diagrams and loop
integrations. Concretely, this expression is at the hadron level,
thus our strategy is to derive all the effective operators at the
quark level by carrying out the loop integrations and then in
terms of the wavefunctions of the mesons, we obtain the decay
amplitude at the hadron level, indeed our goal is to derive
$g_{eff}$ in eq.(2).

Without the weak-binding approximation which was adopted in
literature\cite{Korner,coupling}, and keeping the masses of the
light quarks at the propagators, we re-formulate the amplitude.
The amplitude of $V\rightarrow P\gamma$ can be divided into three
pieces which correspond to Fig. 2 (a), (b) and (c) respectively.

For Fig. 2(a), we have
\begin{eqnarray}
M_{A}  &=& \frac{i \pi^{2}eQ_{Q}g_{s}^{4}T^{a}T^{b} \otimes
T^{b}T^{a}}{(({1\over 2}p_{_V}-q)^{2}-m_{Q}^{2})(2\pi)^{4}}
\varepsilon^{*\mu}(\gamma) \nonumber \\
&&[D^{\rho\nu}_1(x,m_q)({1\over
2}p_{_V}-q)^{\sigma}O^{\mathbf{1}}_{\mu\sigma\rho\nu}+
D^{\rho\nu}_1(x,m_q)m_{Q}O^{\mathbf{2}}_{\mu\rho\nu}+
D^{\nu}_2(x,m_q)({1\over
2}p_{_V}-q)^{\sigma}m_{Q}O^{\mathbf{3}}_{\mu\nu\sigma}\nonumber
\\
&&+
D^{\nu}_2(x,m_q)m_{Q}^{2}O^{\mathbf{4}}_{\mu\nu}],\nonumber \\
\end{eqnarray}
where
\begin{eqnarray}
&&O^{\mathbf{1}}_{\mu\sigma\rho\nu}=-2ig_{\rho\nu}\varepsilon_{\beta\sigma\mu
a}\bar{\upsilon}_{Q}\gamma^{a}u_{Q}\bar{u_{q}}\gamma^{\beta}\gamma_{5}\upsilon_{q}+
2i\varepsilon_{\nu\sigma\mu\theta}\bar{\upsilon}_{Q}\gamma^{\theta}u_{Q}
\bar{u_{q}}\gamma_{\rho}\gamma_{5}\upsilon_{q},\nonumber\\
&&O^{\mathbf{2}}_{\mu\rho\nu}=-2g_{\alpha\mu}g_{a\nu}\varepsilon^{\beta
a\alpha
b}\bar{\upsilon}_{Q}\sigma_{\rho\beta}u_{Q}\bar{u_{q}}\gamma_{b}\gamma_{5}\upsilon_{q}+g_{\rho\mu}g_{a\nu}\varepsilon^{\beta
a\alpha
b}\bar{\upsilon}_{Q}\sigma_{\rho\beta}u_{Q}\bar{u_{q}}\gamma_{b}
\gamma_{5}\upsilon_{q}, \nonumber\\
&&O^{\mathbf{3}}_{\mu\nu\sigma}=-2g_{\alpha\sigma}g_{a\nu}\varepsilon^{\beta
a\alpha
b}\bar{\upsilon}_{Q}\sigma_{\beta\mu}u_{Q}\bar{u_{q}}\gamma_{b}\gamma_{5}\upsilon_{q}
-2g_{\beta\mu}g_{a\nu}\varepsilon^{\beta a\alpha
b}\bar{\upsilon}_{Q}\sigma_{\alpha\sigma}u_{Q}\bar{u_{q}}\gamma_{b}\gamma_{5}\upsilon_{q}
\nonumber \\
&&+g_{\sigma\mu}g_{a\nu}\varepsilon^{\beta a\alpha
b}\bar{\upsilon}_{Q}\sigma_{\alpha\beta}u_{Q}\bar{u_{q}}\gamma_{b}
\gamma_{5}\upsilon_{q}, \nonumber\\
&&O^{\mathbf{4}}_{\mu\nu}=-2ig_{\beta\mu}g_{a\nu}\varepsilon^{\beta
a\alpha
b}\bar{\upsilon}_{Q}\gamma_{\alpha}u_{Q}\bar{u_{q}}\gamma_{b}\gamma_{5}\upsilon_{q}.
\end{eqnarray}
The functions $D^{\rho\nu}_1(x,m_q),\;D^{\nu}_2(x,m_q)$ are
integrals of four-point functions over the internal momentum $k$
whose explicit forms are given in appendix A.

For Fig.2 (b), the amplitude reads
\begin{eqnarray}
M_{B} & = &\frac{i \pi^{2}eQ_{Q}g_{s}^{4}T^{a}T^{b} \otimes
T^{b}T^{a}}{[(\frac{1}{2}p_{V}-q)^{2}-m_{Q}^{2}](2\pi)^{4}}
\varepsilon^{*\mu}(\gamma)\nonumber  \\ \nonumber
&&[D^{\rho\nu}_1(x,m_q)({1\over
2}p_{_V}-q)^{\sigma}O^{\mathbf{1}}_{\mu\sigma\rho\nu}-D^{\rho\nu}_1(x,m_q)m_{Q}O^{\mathbf{2}}_{\mu\rho\nu}-
D^{\nu}_2(x,m_q)({1\over
2}p_{_V}-q)^{\sigma}m_{Q}O^{\mathbf{3}}_{\mu\nu\sigma}\nonumber
\\
&&+
D^{\nu}_2(x,m_q)m_{Q}^{2}O^{\mathbf{4}}_{\mu\nu}],\nonumber \\
\end{eqnarray}
and the amplitude corresponding to Fig.2 (c) is
\begin{eqnarray}
M_{C} &=& \frac{eQ_{Q}g_{s}^{4}T^{a}T^{b} \otimes T^{b}T^{a}i
\pi^{2}}{(2\pi)^{4}} \varepsilon^{*\mu}(\gamma)
[E^{\nu\rho\sigma}_1(x,m_q)O^{\mathbf{5}}_{\mu\nu\rho\sigma}+
E^{\nu\sigma}_2(x,m_q)m_{Q}O^{\mathbf{6}}_{\mu\nu\sigma}+
E^{\rho\sigma}_3(x,m_q)m_{Q}O^{\mathbf{7}}_{\mu\rho\sigma}],\nonumber \\
\end{eqnarray}
where
\begin{eqnarray}
&&O^{\mathbf{5}}_{\mu\nu\rho\sigma}=i\varepsilon_{\nu\mu\rho\sigma}
\bar{\upsilon}_{Q}\gamma_{\beta}u_{Q}\bar{u_{q}}\gamma^{\beta}\gamma_{5}\upsilon_{q}
-2i\varepsilon_{\nu\mu\rho\theta}\bar{\upsilon}_{Q}
\gamma_{\sigma}u_{Q}\bar{u_{q}}\gamma^{\theta}\gamma_{5}\upsilon_{q},\nonumber
\\
&&O^{\mathbf{6}}_{\mu\nu\sigma}=g_{c\sigma}g_{\mu\nu}\varepsilon^{\beta
c\alpha\theta}\bar{\upsilon}_{Q}\sigma_{\alpha\beta}u_{Q}
\bar{u_{q}}\gamma_{\theta}\gamma_{5}\upsilon_{q},\nonumber\\
&&O^{\mathbf{7}}_{\mu\rho\sigma}=g_{c\sigma}g_{\mu\rho}\varepsilon^{\beta
c\alpha\theta}\bar{\upsilon}_{Q}\sigma_{\alpha\beta}
u_{Q}\bar{u_{q}}\gamma_{\theta}\gamma_{5}\upsilon_{q}.
\end{eqnarray}
The functions
$E^{\nu\rho\sigma}_1(x,m_q),\;E^{\nu\sigma}_2(x,m_q),\;E^{\rho\sigma}_3(x,m_q)$
are three independent integrals of five-point functions over the
internal momentum $k$ and their explicit expressions are also
collected in appendix A.

The contributions of the other three diagrams Fig. 2 (d,e,f) are
similar to that of the first three (a,b,c), to save space, we omit
their expressions in the context.

It is noted that the contribution of Fig.2 (c) is a five-point
Green's function, namely there are five propagators in the loop.
As long as the weak-binding approximation is adopted, the
five-point functions $E_1^{\nu\rho\sigma}(x,m_q)$,
$E_2^{\nu\sigma}(x,m_q)$, $E_3^{\rho\sigma}(x,m_q)$ can be
decomposed into sums of two-point and three-point functions, so
that the calculations are much simplified and analytical
expressions are eventually derived \cite{Korner,Yang}. By
contrary, without the weak-binding approximation, such
decomposition is impossible, unfortunately. Following Denner and
Dittmaier \cite{loop}, we decompose the five-point functions into
a sum of several four-point functions which cannot be integrated
out analytically. Instead, we will calculate the integrals
numerically. Some details about the integration are presented in
Appendix B.

(ii) For $V\rightarrow \rho^0\gamma$.

Serving as a check, we also calculate the branching ratio of
$J/\psi\rightarrow \rho^0\gamma$ in comparison with that of
$J/\psi\rightarrow \pi^0\gamma$. The Feynman diagrams are the same
as that for $J/\psi\rightarrow \pi^0\gamma$, and the explicit
expressions of the amplitudes are given in Appendix A for saving
space.\\

\subsection{The hadronic matrix elements}

With the Feynman diagrams of Fig.2 we derive an effective
lagrangian at the quark level. To obtain the hadronic matrix
elements and then finally the decay rates, one has to evaluate the
hadronic matrix elements. It is well known that the hadronization
happens at the energy scale of $\Lambda_{QCD}$ which is in the
range of non-perturbative QCD, so far, there is no any reliable
way to evaluate the hadronic matrix elements. To do the job, we
need to invoke concrete models. Since the produced meson is
relatively light and its three-momentum is larger than
$\Lambda_{QCD}$, the light-cone wavefunctions seem  to be
plausible for description of the light
mesons\cite{light-cone2,light-cone1,light-cone3}.

The matrix elements are
 \begin{eqnarray}
 &&\left<P|\bar{q}\gamma_{\alpha}\gamma_{5}q \sum_i(C_{i}(x,m_q)O_{i})\bar{Q}
 \gamma_{\beta}Q|V\right>\nonumber\\
&&=-if_{_P}p_{_P\alpha}\int_{0}^{1}dx\phi_{_P}(\mu,x)\sum_i(C_{i}(x,m_q)O_{i})
if_{_V}\varepsilon_{_V\beta}M_{_V},\nonumber\\
 &&\left<P|\bar{q}\gamma_\alpha\gamma_{5}q \sum_i(C_{i}(x,m_q)O_{i})
 \bar{Q}{\sigma_{\beta\rho}}Q|V\right>\nonumber\\
 \nonumber
 &&=-if_{_P}p_{P\alpha}\int_{0}^{1}dx\phi_{_P}(\mu,x)\sum_i(C_{i}(x,m_q)O_{i})if_{_V}
i(\varepsilon_{_V\beta}p_{_V\rho}-p_{_V\beta}\varepsilon_{_V\rho}),
 \end{eqnarray}
where $O_i$'s and $C_i$'s are the operators  and their
coefficients derived in last sub-section. For the pseudoscalars,
the SU(3) flavor wavefuctions are
$$\pi^{0}={u\bar{u}-d\bar{d}\over \sqrt{2}},\;
\eta_{0}={d\bar{d}+u\bar{u}+s\bar{s}\over \sqrt{3}}\;{\rm and}\;
\eta_{8}={d\bar{d}+u\bar{u}-2s\bar{s}\over \sqrt{6}},$$ $\eta$ and
$\eta^{'}$ are mixutures of $\eta_{0}$ and $\eta_{8}$£¬
$$\eta=\cos\theta\eta_{8}-\sin\theta\eta_{0},\;\;\;\;
\eta^{'}=\sin\theta\eta_{8}+\cos\theta\eta_{0}.$$

The normalization of the light-cone wavefunction is defined as
$$\int_{0}^{1}dx\phi(\mu,x)=1.$$
The explicit forms of the light-cone wavefunctions of the light
mesons can be different. In our later calculations, we take three
different types of wavefunctions which are given in
literatures\cite{wave1,wave2,light-cone2,wave3} as
\begin{eqnarray}
&&\label{fun1} \phi_1(\mu,x) = 6x(1-x), \\
&&\label{fun2} \phi_2(\mu,x) = 30x^2(1-x)^2, \\
&&\label{fun3} \phi_3(\mu,x) = {15\over 2}(1-2x)^2[1-(1-2x)^2].
\end{eqnarray}

Finally we obtain the hadronic matrix elements of $M_A,\;M_B$ and
$M_C$ as following
\begin{eqnarray}
&&\langle P\gamma|M_{A}|V\rangle =i\pi^2eQ_Qg_s^4T^aT^b\otimes
T^bT^a{1 \over (2\pi)^4}{1\over ({1\over
2}p_{_V}-q)^2-m_Q^2}{1\over
N_c^2}\varepsilon^{*\mu}(\gamma)\int_0^1dx\phi_P(\mu,x)\nonumber
\\
&&\{f_P f_{V}M_{V}[-2ig_{\rho\nu}\epsilon_{\beta\sigma\mu
a}p_{_P}^\beta
\varepsilon_{_V}^a+2i\epsilon_{\nu\sigma\mu\theta}p_{_P\rho}\varepsilon_{_V}^\theta]\sum_{q=u,d,s}(D^{\rho\nu}_1(x,m_q))
({1\over 2}p_{_V}-q)^\sigma \nonumber \\
&&+if_P f_{V}[-2g_{\alpha\mu}g_{a\nu}\epsilon^{\beta a\alpha
b}p_{_P b}(\varepsilon_{_V \rho}p_{_V
\beta}-\varepsilon_{_V \beta}p_{_V\rho})\nonumber \\
&&+g_{\rho\mu}g_{a\nu}\epsilon^{\beta a\alpha b}p_{_P
b}(\varepsilon_{_V \alpha}p_{_V \beta}-\varepsilon_{_V \beta}p_{_V
\alpha})]\sum_{q=u,d,s}(D^{\rho\nu}_1(x,m_q))m_Q
\nonumber \\
&&+if_P f_{V}[-2g_{\alpha\sigma}g_{a\nu}\epsilon^{\beta a\alpha
b}p_{_P b}(\varepsilon_{_V \beta}p_{_V
\mu}-\varepsilon_{_V \mu}p_{_V \beta})\nonumber \\
&&-2g_{\beta\mu}g_{a\nu}\epsilon^{\beta a\alpha b}p_{_P
b}(\varepsilon_{_V \alpha}p_{_V
\sigma}-\varepsilon_{_V \sigma}p_{_V \alpha})\nonumber \\
&&+g_{\sigma\mu}g_{a\nu}\epsilon^{\beta a\alpha b}p_{_P
b}(\varepsilon_{_V \alpha}p_{_V
\beta}-\varepsilon_{_V \beta}p_{_V \alpha})]\sum_{q=u,d,s}(D^{\nu}_2(x,m_q))({1\over 2}p_{_V}-q)^\sigma m_Q\nonumber \\
&&+f_P f_{V}M_{V}[-2ig_{\beta\mu}g_{a\nu}\epsilon^{\beta a\alpha
b}p_{_P b}\varepsilon_{_V
\alpha}]\sum_{q=u,d,s}(D^{\nu}_2(x,m_q))m_Q^2\},
\end{eqnarray}
\begin{eqnarray}
&&\langle P\gamma|M_{B}|V\rangle =i\pi^2eQ_Qg_s^4T^aT^b\otimes
T^bT^a{1 \over (2\pi)^4}{1\over ({1\over
2}p_{_V}-q)^2-m_Q^2}{1\over
N_c^2}\varepsilon^{*\mu}(\gamma)\int_0^1dx\phi_P(\mu,x)\nonumber
\\
&&\{f_P f_{V}M_{V}[-2ig_{\rho\nu}\epsilon_{\beta\sigma\mu
a}p_{_P}^\beta
\varepsilon_{_V}^a+2i\epsilon_{\nu\sigma\mu\theta}p_{_P\rho}\varepsilon_{_V}^\theta]\sum_{q=u,d,s}(D^{\rho\nu}_1(x,m_q))
({1\over 2}p_{_V}-q)^\sigma \nonumber \\
&&-if_P f_{V}[-2g_{\alpha\mu}g_{a\nu}\epsilon^{\beta a\alpha
b}p_{_P b}(\varepsilon_{_V \rho}p_{_V
\beta}-\varepsilon_{_V \beta}p_{_V\rho})\nonumber \\
&&+g_{\rho\mu}g_{a\nu}\epsilon^{\beta a\alpha b}p_{_P
b}(\varepsilon_{_V \alpha}p_{_V \beta}-\varepsilon_{_V \beta}p_{_V
\alpha})]\sum_{q=u,d,s}(D^{\rho\nu}_1(x,m_q))m_Q
\nonumber \\
&&-if_P f_{V}[-2g_{\alpha\sigma}g_{a\nu}\epsilon^{\beta a\alpha
b}p_{_P b}(\varepsilon_{_V \beta}p_{_V
\mu}-\varepsilon_{_V \mu}p_{_V \beta})\nonumber \\
&&-2g_{\beta\mu}g_{a\nu}\epsilon^{\beta a\alpha b}p_{_P
b}(\varepsilon_{_V \alpha}p_{_V
\sigma}-\varepsilon_{_V \sigma}p_{_V \alpha})\nonumber \\
&&+g_{\sigma\mu}g_{a\nu}\epsilon^{\beta a\alpha b}p_{_P
b}(\varepsilon_{_V \alpha}p_{_V
\beta}-\varepsilon_{_V \beta}p_{_V \alpha})]\sum_{q=u,d,s}(D^{\nu}_2(x,m_q))({1\over 2}p_{_V}-q)^\sigma m_Q\nonumber \\
&&+f_P f_{V}M_{V}[-2ig_{\beta\mu}g_{a\nu}\epsilon^{\beta a\alpha
b}p_{_P b}\varepsilon_{_V
\alpha}]\sum_{q=u,d,s}(D^{\nu}_2(x,m_q))m_Q^2\},
\end{eqnarray}
and
\begin{eqnarray}
&&\langle P\gamma|M_{C}|V \rangle =i\pi^2eQ_Qg_s^4T^aT^b\otimes
T^bT^a{1 \over (2\pi)^4}{1\over
N_c^2}\varepsilon^{*\mu}(\gamma)\int_0^1dx\phi_P(\mu,x)\nonumber
\\
&&\{f_P f_{V}M_{V}[i\epsilon_{\nu\mu\rho\sigma}p_{_P}^\beta
\varepsilon_{_V\beta}-2i\epsilon_{\nu\mu\rho\theta}p_{_P}^\theta\varepsilon_{_V\sigma}]
\sum_{q=u,d,s}(E^{\nu\rho\sigma}_1(x,m_q))\nonumber
\\
&&+if_P f_{V}[g_{\sigma c}\epsilon^{\beta
c\tau\theta}g_{\mu\nu}p_{_P\theta}(\varepsilon_{_V\tau}p_{_V\beta}-\varepsilon_{_V\beta}p_{_V\tau})]
\sum_{q=u,d,s}(E^{\nu\sigma}_2(x,m_q))m_Q\nonumber \\
&&+if_P f_{V}[g_{\sigma c}\epsilon^{\beta
c\tau\theta}g_{\mu\rho}p_{_P\theta}(\varepsilon_{_V\tau}p_{_V\beta}-\varepsilon_{_V\beta}p_{_V\tau})]
\sum_{q=u,d,s}(E^{\rho\sigma}_3(x,m_q))m_Q\}.
\end{eqnarray}

\section{Numerical results}

In this section, we present our numerical results.

In the numerical computations, there is a mild Infrared (IR)
divergence problem. Namely, when we carry out the loop integration
and convolution integrals of the effective operators with the
light-cone wavefunction of the produced meson, an IR-divergence
emerges, but it is not as serious as that in the B-meson decays
and can be removed in simple ways. Our strategy to deal with the
IR problem is standard, namely we assign a small mass to the gluon
and vary it to check if the result is stable. Practically, we set
the small mass to be from $10^{-4}$ MeV to $10^{-6}$ MeV and find
that the result has only negligible changes. Therefore we can
trust the obtained result which is free of the IR problem. Our
final results given in all the following tables correspond to the
gluon mass of $10^{-5}$ MeV.

\subsection{Input parameters}
The input parameters which we are going to use in the numerical
computations are taken as
follows\cite{data0,coupling,light-cone1,data1,data2}:$f_{J/\psi} =
551MeV$, $f_\Upsilon = 710MeV$, $f_\pi = 131MeV$, $f_\eta =
f_{\eta^{'}} = 157MeV$, $f_\rho = 198MeV$, $M_{J/\psi} =
3096.87MeV$, $M_\Upsilon = 9460.3MeV$, $m_\pi = 134.9766MeV$,
$m_\eta = 547.75MeV$, $m_{\eta^{'}} = 957.78MeV$, $m_\rho =
775.8MeV$, $\alpha_s(m_c) = 0.26$, $\alpha_s(m_b) = 0.17$, $m_c =
1300MeV$, $m_b = 4700MeV$, the mixing angle $\theta = -11^\circ$,
and three possible distribution amplitudes of pseudoscalar meson
are given in Eqs.(\ref{fun1},\ref{fun2},\ref{fun3}).

We will present the the resultant decay rates corresponding to the
three different distribution amplitudes respectively in the
following tables.

\subsection{Numerical results of process $V\to P \gamma$}
With the above parameters, the theoretical values for the decay
width of these processes in the rest frame of $J/\psi$, are shown
in Table 1.
\pagebreak[4]
\begin{table}[h]
\caption{The decay branching ratio of $J/\psi \to \pi^0 \gamma$,
$J/\psi \to \eta \gamma$ and $J/\psi \to \eta^{'} \gamma$ in the
rest frame of $J/\psi$ and the three columns correspond to the
three different parton distribution amplitudes of the produced
pseudoscalar mesons ($\pi^0,\;\eta,\;\eta'$).}
\begin{center}
\begin{tabular}{|c|c|c|c|c|c|c|c|} \hline
Processes &$m_u$&$m_d$&$m_s$&$10^5\times BR(\phi_1)$&$10^5\times BR(\phi_2)$&$10^5\times BR(\phi_3)$&Experimental data\\
\hline
& 1.5 & 4 & 0 & 2.71325 & 0.402025 & 25.9819 & \\
\cline{2-7}
& 2 & 4 & 0 & 0.895075 & 0.292837 & 9.09646 & \\
\cline{2-7}
$J/\psi \to \pi^0 \gamma$ & 3 & 5 & 0 & 1.27478 & 0.428238 & 11.8012 & $(3.9\pm 1.3)\times10^{-5}$\\
\cline{2-7}
& 3 & 7 & 0 & 6.43086 & 2.21318 & 55.5486 & \\
\cline{2-7}
& 4 & 6 & 0 & 1.64827 & 0.565255 & 14.1993 & \\
\hline
&$m_u$&$m_d$&$m_s$&$10^4\times BR(\phi_1)$&$10^4\times BR(\phi_2)$&$10^4\times BR(\phi_3)$&\\
\cline{2-7}
& 2 & 4 & 80 & 2.41823 & 1.68707 & 7.23647 & \\
\cline{2-7}
& 2 & 5 & 90 & 2.46188 & 1.69669 & 7.44466 &\\
\cline{2-7}
$J/\psi \to \eta \gamma$ & 3 & 5 & 100 & 2.56809 & 1.72051 & 8.12141 & $(8.6\pm 0.8)\times10^{-4}$\\
\cline{2-7}
& 2 & 6 & 110 & 2.48921 & 1.6743 & 7.97546 & \\
\cline{2-7}
& 4 & 6 & 120 & 2.50536 & 1.66464 & 8.12365 & \\
\cline{2-7}
& 3 & 7 & 130 & 2.55453 & 1.65799 & 8.6264 & \\
\hline
&$m_u$&$m_d$&$m_s$&$10^3\times BR(\phi_1)$&$10^3\times BR(\phi_2)$&$10^3\times BR(\phi_3)$&\\
\cline{2-7}
& 2 & 4 & 80 & 1.19297 & 1.07802 & 1.92765 &\\
\cline{2-7}
& 2 & 5 & 90 & 1.18068 & 1.07066 & 1.90548 &\\
\cline{2-7}
$J/\psi \to \eta^{'} \gamma$ & 3 & 5 & 100 & 1.17033 & 1.06154 & 1.87308 & $(4.3\pm 0.3)\times 10^{-3}$\\
\cline{2-7}
& 2 & 6 & 110 & 1.15948 & 1.05165 & 1.84638 & \\
\cline{2-7}
& 4 & 6 & 120 & 1.14416 & 1.03956 & 1.8022 & \\
\cline{2-7}
& 3 & 7 & 130 & 1.12954 & 1.02747 & 1.76655 & \\
\hline
\end{tabular}
\end{center}
\end{table}

Obviously, the same procedure can be applied to the radiative
decays of $\Upsilon\rightarrow \pi^0(\eta,\eta')\gamma$. We
calculate the widths with the same parameters by replacing c-quark
in $J/\psi$ by b-quark in $\Upsilon$, $Q_c$ by $Q_b$ and $m_c$ by
$m_b$. Then we obtain the numerical results which are shown in
Table 2. \pagebreak[4]
\begin{table}[h]
\caption{The decay branching ratio of $\Upsilon \to \pi^0 \gamma$,
$\Upsilon \to \eta \gamma$ and $\Upsilon \to \eta^{'} \gamma$ in
the rest frame of $\Upsilon$ and the three columns correspond to
the three different distribution amplitudes of the produced
pseudoscalar mesons.}
\begin{center}
\begin{tabular}{|c|c|c|c|c|c|c|c|} \hline
Processes &$m_u$&$m_d$&$m_s$&$10^{5}\times BR(\phi_1)$&$10^{5}\times BR(\phi_2)$&$10^{5}\times BR(\phi_3)$&Experimental data\\
\hline
& 1.5 & 4 & 0 & 0.958428 & 0.674172 & 2.78981 & \\
\cline{2-7}
& 2 & 4 & 0 & 0.634743 & 0.478474 & 1.60618 & \\
\cline{2-7}
$\Upsilon \to \pi^0 \gamma$ & 3 & 5 & 0 & 0.575449 & 0.618915 & 0.445925 & $<3\times10^{-5}$\\
\cline{2-7}
& 3 & 7 & 0 & 1.92777 & 2.81325 & 0.5723 & \\
\cline{2-7}
& 4 & 6 & 0 & 0.4889 & 0.723972 & 0.071549 & \\
\hline
&$m_u$&$m_d$&$m_s$&$10^{5}\times BR(\phi_1)$&$10^{5}\times BR(\phi_2)$&$10^5\times BR(\phi_3)$&\\
\cline{2-7}
& 2 & 4 & 80 & 0.264287 & 0.037964 & 9.31202 & \\
\cline{2-7}
& 2 & 5 & 90 & 0.467742 & 0.0426104 & 13.9632 & \\
\cline{2-7}
$\Upsilon \to \eta \gamma$ & 3 & 5 & 100 & 0.772832 & 0.0571983 & 19.97 & $<2.1\times10^{-5}$\\
\cline{2-7}
& 2 & 6 & 110 & 1.20528 & 0.0916147 & 27.4375 & \\
\cline{2-7}
& 4 & 6 & 120 & 1.80313 & 0.158933 & 36.6147 & \\
\cline{2-7}
& 3 & 7 & 130 & 2.59821 & 0.276932 & 47.4917 & \\
\hline
&$m_u$&$m_d$&$m_s$&$10^4\times BR(\phi_1)$&$10^4\times BR(\phi_2)$&$10^4\times BR(\phi_3)$&\\
\cline{2-7}
& 2 & 4 & 80 & 9.58491 & 8.54894 & 14.7284 & \\
\cline{2-7}
& 2 & 5 & 90 & 9.57026 & 8.55094 & 14.6 & \\
\cline{2-7} $\Upsilon \to \eta^{'} \gamma$ & 3 & 5 & 100 & 9.54632 & 8.5466 & 14.4558 & $<1.6\times10^{-5}$\\
\cline{2-7}
& 2 & 6 & 110 & 9.51302 & 7.99717 & 14.2925 & \\
\cline{2-7}
& 4 & 6 & 120 & 9.47596 & 8.51945 & 14.1333 & \\
\cline{2-7}
& 3 & 7 & 130 & 9.42768 & 8.49574 & 13.9498 & \\
\hline
\end{tabular}
\end{center}
\end{table}
\\
As discussed above, as a check, we evaluate the
decay width of $J/\psi\rightarrow \rho^0\gamma$ which corresponds
to an effective three-vector vertex and has not been observed yet.
The results are tabulated in Table 3.

\begin{table}[htb]
\caption{The decay branching ratio of $J/\psi \to \rho^0 \gamma$
in the rest frame of $J/\psi$, and the three columns correspond to
the three different distribution amplitudes of $\rho^0$.}
\begin{center}
\begin{tabular}{|c|c|c|c|c|c|c|c|}
  \hline
Processes &$m_u$&$m_d$&$m_s$&$10^{13}\times BR(\phi_1)$&$10^{13}\times BR(\phi_2)$&$10^{13}\times BR(\phi_3)$&Experimental data\\
  \hline
$J/\psi \to \rho^0 \gamma$ & 4 & 6 & 0 & 4.74477 & 2.07037 & 37.9532 & $-$\\
  \hline
\end{tabular}
\end{center}
\end{table}
\pagebreak[4]
\section{Conclusion and Discussion}

In this work, we re-study the OZI forbidden radiative decays of
othoquarkonia in the framework of perturbative QCD. In the
process, we do not take the weak-binding approximation or set the
light quark mass to be zero and carry out a complete integration
of the five- and four-point functions. In this scenario, we can
take into account the SU(3) and isospin violation which would
result in non-zero rates for $J/\psi\rightarrow \pi^0\gamma$ and
$J/\psi\rightarrow \eta_8\gamma$. However, on the other side,
without such approximation, we cannot derive elegant analytical
expressions for the amplitudes as done by K\"orner et al.
\cite{Korner} and Yang \cite{Yang}. Instead, we need to invoke
complicated computer programs and we follow the recipe given by
Denner and Dittmaier \cite{loop} to reduce the five-point
functions into a sum of four-point functions and then use the
developed computer program "LoopTools" \cite{program} to carry out
the integrations.

Another difficult point is to evaluate the hadronic matrix
elements which are fully governed by the non-perturbative QCD. The
initial heavy quarkonium ($J/\psi$ or $\Upsilon$) are composed of
only heavy quarks, so that the on-shell approximation is
reasonable and its contribution to the amplitude can be described
by its wavefunction at origin, i.e. $R_V(0)$
\cite{Korner,coupling}. As we abandon the weak-binding
approximation for the produced light meson, the simple description
in \cite{Korner} for the produced meson by
$R_{PS}(0)$\cite{Korner,coupling} can no longer be adopted. Since
the produced pseudoscalar meson is light, the light-cone
wavefunctions seem to be applicable for the calculations. There
are several typical different light-cone distribution amplitudes
for the light mesons, so far, one cannot determine which one is
the most suitable. Thus we adopt all the three for our
calculations and the corresponding results are listed in the
tables of last section.

Our results for $J/\psi\rightarrow P\gamma$ where $P$ stands for
$\pi^0,\;\eta,\;\eta'$, are quantitatively in agreement with the
experimental measurements. For $J/\psi\rightarrow \pi^0\gamma$
which is an isospin violating process because it is proportional
to an asymmetry of u and d quarks, within reasonable ranges of the
masses of u and d quarks, all the three distribution amplitudes
can result in values in agreement with data. For
$J/\psi\rightarrow \eta\gamma$, it seems that only $\phi_3$ can
give the values in good agreement with data. For
$J/\psi\rightarrow \eta'\gamma$, all the obtained values are
slightly smaller than the data. This small declination can be
understood, because theoretically the results depend on the masses
of light quarks and the QCD coupling, (here we do not include the
running of $\alpha_s$ as in \cite{Korner}), and experimentally,
all the concerned widths are small and certain measurement errors
are unavoidable. We can expect that the CLEO and BES III which
will be running in 2007, can make more precise measurements to
testify the results. Moreover, we calculate the decay width of
$J/\psi \rightarrow \rho^0\gamma$ which has not been measured yet.
This result will be measured in near future and the data can
provide some information about our understanding of both
perturbative and non-perturbative QCD (the ansatz for light-cone
wavefunctions of mesons).

As we turn to the radiative decays of $\Upsilon$, the situation
seems peculiar. It is noted that $\Upsilon\rightarrow \pi^0\gamma$
is related to isospin violation, $\Upsilon\rightarrow \eta\gamma$
is related to SU(3) violation because $\eta$ has a large fraction
of $\eta_8$, only $\Upsilon\rightarrow \eta'\gamma$ conserves
SU(3), as $\eta'$ contains mainly $\eta_0$. Thus one can expect
$\Gamma(\Upsilon\rightarrow \pi^0\gamma)<
\Gamma(\Upsilon\rightarrow \eta\gamma)<\Gamma( \Upsilon\rightarrow
\eta'\gamma)$. For radiative decays of $J/\psi$, this sequence
obviously holds and our calculations confirm this pattern.
However, for $\Upsilon$, the measurements seem not to follow the
pattern \cite{data0}. Of course the data only set upper bounds on
these decay modes, there is still some possibility to upset this
pattern.

Our theoretical results for $\Upsilon$ still follow the sequence,
and for $\Upsilon\rightarrow \pi^0\gamma$ and $\Upsilon\rightarrow
\eta\gamma$, the values are consistent with the experimental upper
bonds, but for $\Upsilon\rightarrow \eta'\gamma$, the calculated
value is obviously larger than the upper bound set by the present
experimental measurement. It may indicate some anomaly in these
decay modes of $\Upsilon$ and the reason is worth further studies
both theoretically and experimentally.

The anomaly problem has been observed by several authors
\cite{Yang,Novikov,Ma,Yellow}, all agree that in the perturbative
framework which is the main content of this work, this anomaly
exists. To reconcile the theoretical results with the data, the
gluon contents in $\eta$ and $\eta'$ are considered
\cite{Novikov,Ma,Yellow}, and the non-perturbative matrix elements
$<0|G\tilde G|\eta^{(')}>$ was evaluated by Feldman and Kroll
\cite{Novikov}. It seems that the non-perturbative effects
alleviate the discrepancy between the theoretical results and
data. However, as pointed in \cite{Yellow}, the problem is not
fully solved yet and needs to be investigated further. In our
work, we are only dealing with the perturbative part and confirm
existence of the anomaly as discovered before. We will investigate
the anomaly and the probable non-perturbative effects in a wider
range. Moreover, in the works\cite{Yang,Novikov,Ma,Yellow}, the
$\gamma\pi^0$ case was dismissed because of its smallness, by
contraries, we explicitly keep the light-quark masses and obtain
the corresponding results for $\gamma\pi^0$ final state in the
radiative decays of both $J/\psi$ and $\Upsilon$.

Our starting point is indeed the same as that of K\"orner et al.
\cite{Korner} and Yang \cite{Yang}, except we do not take any
approximation and carry out the full integration of the five- and
four-point functions. We find that the numerical results for
$J/\psi$ are consistent with data, but there is an anomaly for
$\Upsilon\rightarrow\eta'\gamma$ if the present measurement is
correct. All these need
more and careful investigations.\\

\noindent Acknowledgement:

This work is partly supported by the National Natural Science
Foundation of China. We benefit greatly from discussions with Dr.
Y.D. Yang and he kindly introduced his work on this subject to us.
We are very grateful to J.P. Ma for his encouragement and fruitful
discussions.

\noindent{ Appendix A}\\

The integrations corresponding to Fig.2 (a), (b) and (c) are
\begin{eqnarray}
M_{A} && = \int\frac{d^4k}{(2\pi)^4} \varepsilon^{*
\mu}(\gamma)\bar{v_{Q}}(p_{2})(-ig_sT^{a}\gamma_{\alpha}){i\over
\rlap /p_1-\rlap /q+\rlap /k-m_Q}(-ig_sT^{b}\gamma_{\beta}){i\over
\rlap /p_1-\rlap /q-m_Q}\nonumber
\\ &&(-ieQ_{Q}\gamma_{\mu})u_{Q}(p_{1})
\bar{u_{q}}(p_{3})(-ig_sT^{b}\gamma^{\beta}){i\over \rlap
/p_3+\rlap
/k-m_q}(-ig_sT^{a}\gamma^{\alpha})v_{q}(p_{4})\nonumber\\
&&{-i\over (p_1+p_2+k-q)^2}\frac{-i}{k^2} \nonumber \\ \nonumber
&& = \int\frac{d^4k}{(2\pi)^4}eQ_Q g_s^4 T^aT^b\otimes T^bT^a
\bar{v_{Q}}(p_{2})\gamma_{\alpha}(\rlap /p_1-\rlap /q+\rlap
/k+m_Q)\gamma_{\beta}(\rlap /p_1-\rlap
/q+m_Q)\gamma_{\mu}u_{Q}(p_{1}) \nonumber
\\ \nonumber &&
\bar{u_{q}}(p_{3})\gamma^{\beta}(\rlap /p_3+\rlap
/k+m_q)\gamma^{\alpha}v_{q}(p_{4})\varepsilon^{* \mu}(\gamma)
\nonumber \\ && \frac{1}{k^2(p_1+p_2+k-q)^2
[(p_1-q+k)^2-m_Q^2][(p_1-q)^2-m_Q^2][(p_3+k)^2-m_q^2]},
\;\;\;\;\;(A1),\nonumber
\end{eqnarray}
\begin{eqnarray}
M_{B} && = \int\frac{d^4k}{(2\pi)^4} \varepsilon^{*
\mu}(\gamma)\bar{v_{Q}}(p_{2})(-ieQ_{Q}\gamma_{\mu}){i\over \rlap
/q-\rlap /p_2-m_Q}(-ig_{s}T^{a}\gamma_{\alpha}){i\over \rlap
/p_1+\rlap /k-m_Q}\nonumber
\\ &&(-ig_{s}T^{b}\gamma_{\beta})u_{Q}(p_{1})
\bar{u_{q}}(p_{3})(-ig_{s}T^{b}\gamma^{\beta}){i\over \rlap
/p_3+\rlap
/k-m_q}(-ig_{s}T^{a}\gamma^{\alpha})v_{q}(p_{4})\nonumber\\
&&{-i\over (p_1+p_2+k-q)^2}\frac{-i}{k^2} \nonumber \\ &&
=\int\frac{d^4k}{(2\pi)^4}\varepsilon^{* \mu}(\gamma)eQ_c g_s^4
T^aT^b\otimes T^bT^a \bar{v_{Q}}(p_{2})\gamma_{\mu}(\rlap /q-\rlap
/p_2+m_Q)\gamma_{\alpha}(\rlap /p_1+\rlap
/k+m_Q)\gamma_{\beta}u_{Q}(p_{1}) \nonumber
\\ &&
\bar{u_{q}}(p_{3})\gamma^{\beta}(\rlap /p_3+\rlap
/k+m_q)\gamma^{\alpha}v_{q}(p_{4}) \nonumber \\ &&
\frac{1}{k^2(p_1+p_2+k-q)^2 [(q-p_2)^2-m_Q^2]\nonumber
[(p_1+k)^2-m_Q^2][(p_3+k)^2-m_q^2]},\;\;\;\;\;(A2),\nonumber
\end{eqnarray}
\begin{eqnarray}
M_{C} && = \int\frac{d^4k}{(2\pi)^4} \varepsilon^{*
\mu}(\gamma)\bar{v_{Q}}(p_{2})(-ig_sT^{a}\gamma_{\alpha}){i\over
\rlap /p_1-\rlap /q+\rlap /k-m_Q}(-ieQ_{Q}\gamma_{\mu})\nonumber
\\
&&{i\over \rlap /p_1+\rlap
/k-m_Q}(-ig_sT^{b}\gamma_{\beta})u_{Q}(p_{1}) \nonumber
\\ &&
\bar{u_{q}}(p_{3})(-ig_sT^{b}\gamma^{\beta}){i\over \rlap
/p_3+\rlap
/k-m_q}(-ig_sT^{a}\gamma^{\alpha})v_{q}(p_{4})\nonumber\\
&&\frac{-i}{(p_1+p_2+k-q)^2}\frac{-i}{k^2} \nonumber \\
&& =eQ_Q g_s^4 T^aT^b\otimes T^bT^a \int\frac{d^4k}{(2\pi)^4}
\bar{v_{Q}}(p_{2})\gamma_{\alpha}(\rlap /p_1-\rlap /q+\rlap
/k+m_Q)\gamma_{\mu}(\rlap /p_1+\rlap
/k+m_Q)\gamma_{\beta}u_{Q}(p_{1}) \nonumber
\\ &&
\bar{u_{q}}(p_{3})\gamma^{\beta}(\rlap /p_3+\rlap
/k+m_q)\gamma^{\alpha}v_{q}(p_{4})\varepsilon^{* \mu}(\gamma)
\nonumber \\
&& \frac{1}{k^2(p_1+p_2+k-q)^2
[(p_1-q+k)^2-m_Q^2][(p_1+k)^2-m_Q^2][(p_3+k)^2-m_q^2]},~~~(A3).
\nonumber
\end{eqnarray}

The expressions for the process $J/\psi\rightarrow \rho^0\gamma$
are
\begin{eqnarray}
&&M_{A}
 = \frac{i \pi^{2}eQ_{Q}g_{s}^{4}T^{a}T^{b} \otimes
T^{b}T^{a}}{(({1\over 2} p_{_V}-q)^{2}-m_{Q}^{2})(2\pi)^{4}}
\varepsilon^{*\mu}(\gamma) \nonumber \\ \nonumber
&&[D^{\rho\nu}_1(x,m_q)({1\over 2}
p_{_V}-q)^{\sigma}\overline{O^{\mathbf{4}}_{1\mu\sigma\rho\nu}}+
D^{\rho\nu}_1(x,m_q)m_{Q}\overline{O^{\mathbf{4}}_{2\mu\rho\nu}}+
D^{\nu}_2(x,m_q)({1\over 2}
p_{_V}-q)^{\sigma}m_{Q}\overline{O^{\mathbf{4}}_{3\mu\nu\sigma}}\nonumber
\\
&&+
D^{\nu}_2(x,m_q)m_{Q}^{2}\overline{O^{\mathbf{4}}_{4\mu\nu}}], \nonumber \\
\nonumber
\end{eqnarray}
where
\begin{eqnarray}
&&\overline{O^{\mathbf{1}}_{\mu\sigma\rho\nu}}
=2g_{\nu\sigma}\bar{\upsilon}_{Q}\gamma_\mu
u_{Q}\bar{u_{q}}\gamma_\rho\upsilon_{q}+
2g_{\sigma\mu}\bar{\upsilon}_{Q}\gamma_\nu
u_{Q}\bar{u_{q}}\gamma_{\rho}\upsilon_{q}-
2g_{\mu\nu}\bar{\upsilon}_{Q}\gamma_\sigma
u_{Q}\bar{u_{q}}\gamma_{\rho}\upsilon_{q}\nonumber \\
&&+ 2g_{\rho\nu}\bar{\upsilon}_{Q}\gamma_\mu
u_{Q}\bar{u_{q}}\gamma_{\sigma}\upsilon_{q}+
2g_{\rho\nu}g_{\sigma\mu}\bar{\upsilon}_{Q}\gamma_\beta
u_{Q}\bar{u_{q}}\gamma^{\beta}\upsilon_{q}-
2g_{\rho\nu}\bar{\upsilon}_{Q}\gamma_\sigma u_{Q}\bar{u_{q}}\gamma_{\mu}\upsilon_{q}\nonumber\\
&&\overline{O^{\mathbf{2}}_{\mu\rho\nu}}=2i\bar{\upsilon}_{Q}\sigma_{\nu\mu}u_{Q}\bar{u_{q}}\gamma_{rho}\upsilon_{q}
+2ig_{\rho\nu}\bar{\upsilon}_{Q}\sigma_{\beta\mu}u_{Q}\bar{u_{q}}\gamma^{\beta}\upsilon_{q} \nonumber\\
&&\overline{O^{\mathbf{3}}_{\mu\nu\sigma}}=2i\bar{\upsilon}_{Q}\sigma_{\sigma\mu}u_{Q}\bar{u_{q}}\gamma_{\nu}\upsilon_{q}\nonumber\\
&&\overline{O^{\mathbf{4}}_{\mu\nu}}=2\bar{\upsilon}_{Q}\gamma_{\mu}u_{Q}\bar{u_{q}}\gamma_{\nu}\upsilon_{q}. \nonumber \\
\nonumber
\end{eqnarray}
\begin{eqnarray}
&&M_{B}
 = \frac{i \pi^{2}eQ_{Q}g_{s}^{4}T^{a}T^{b} \otimes
T^{b}T^{a}}{(({1\over 2} p_{_V}-q)^{2}-m_{Q}^{2})(2\pi)^{4}}
\varepsilon^{*\mu}(\gamma) \nonumber \\ \nonumber
&&[-D^{\rho\nu}_1(x,m_q)({1\over 2}
p_{_V}-q)^{\sigma}\overline{O^{\mathbf{1}}_{\mu\sigma\rho\nu}}+
D^{\rho\nu}_1(x,m_q)m_{Q}\overline{O^{\mathbf{2}}_{\mu\rho\nu}}+
D^{\nu}_2(x,m_q)({1\over 2}
p_{_V}-q)^{\sigma}m_{Q}\overline{O^{\mathbf{3}}_{\mu\nu\sigma}}\nonumber
\\
&&-
D^{\nu}_2(x,m_q)m_{Q}^{2}\overline{O^{\mathbf{4}}_{\mu\nu}}]. \nonumber \\
\nonumber
\end{eqnarray}
The corresponding hadronic matrix elements are
 \begin{eqnarray}
 &&\left<V^{'}|\bar{q}\gamma_{\alpha}q \sum_i(C_{i}(x,m_q)\overline{O_{i}})\bar{Q}\gamma_{\beta}Q|V\right>\nonumber\\
&&=if_{_{V^{'}}}\varepsilon_{_{V^{'}\alpha}}m_{_{V^{'}}}\int_{0}^{1}dx\phi_{_{V^{'}}}(\mu,x)\sum_i(C_{i}(x,m_q)\overline{O_{i}})if_{_V}\varepsilon_{_V\beta}M_{_V}\nonumber\\
\nonumber
 \nonumber
 &&\left<V^{'}|\bar{q}\gamma_\alpha q \sum_i(C_{i}(x,m_q)\overline{O_{i}})\bar{Q}{\sigma_{\beta\rho}}Q|V\right>\nonumber\\
 \nonumber
 &&=if_{_{V^{'}}}\varepsilon_{_{V^{'}\alpha}}m_{_{V^{'}}}\int_{0}^{1}dx\phi_{_{V^{'}}}(\mu,x)\sum_i(C_{i}(x,m_q)\overline{O_{i}})if_{_V}
i(\varepsilon_{_V\beta}p_{_V\rho}-p_{_V\beta}\varepsilon_{_V\rho}).\nonumber\\
\nonumber
 \end{eqnarray}
Concretely the hadronic matrix elements of $M_A$ and $M_B$ are
\begin{eqnarray}
&&\langle V^{'}\gamma|M_{A}|V\rangle \nonumber \\
&&=i\pi^2eQ_Qg_s^4T^aT^b\otimes T^bT^a{1 \over (2\pi)^4}{1\over
({1\over 2}p_{_V}-q)^2-m_Q^2}{1\over
N_c^2}\varepsilon^{*\mu}(\gamma)\int_0^1dx\phi_P(\mu,x)\nonumber
\\
&&\{if_{_{V^{'}}}m_{_{V^{'}}}if_{_V}M_{_V}[2g_{\nu\sigma}\varepsilon_{_V\mu}\varepsilon_{_{V^{'}\rho}}
+2g_{\sigma\mu}\varepsilon_{_V\nu}\varepsilon_{_{V^{'}\rho}}-
2g_{\mu\nu}\varepsilon_{_V\sigma}\varepsilon_{_{V^{'}\rho}}+
2g_{\rho\nu}\varepsilon_{_V\mu}\varepsilon_{_{V^{'}\sigma}}\nonumber
\\
&&+
2g_{\rho\nu}g_{\sigma\mu}\varepsilon_{_V\beta}\varepsilon_{_{V^{'}}}^\beta-
2g_{\rho\nu}\varepsilon_{_V\sigma}\varepsilon_{_{V^{'}\mu}}]\sum_{q=u,d,s}(D^{\rho\nu}_1(x,m_q))
({1\over 2}p_{_V}-q)^\sigma \nonumber \\
&&+if_{_{V^{'}}}m_{_{V^{'}}}if_{_V}i[
2i(\varepsilon_{_V\nu}p_{_V\mu}-p_{_V\nu}\varepsilon_{_V\mu})\varepsilon_{_{V^{'}\rho}}
+2ig_{\rho\nu}(\varepsilon_{_V\beta}p_{_V\mu}-p_{_V\beta}\varepsilon_{_V\mu})\varepsilon_{_{V^{'}}}^\beta
]\sum_{q=u,d,s}(D^{\rho\nu}_1(x,m_q))m_Q\nonumber \\
&&+if_{_{V^{'}}}m_{_{V^{'}}}if_{_V}i[2i(\varepsilon_{_V\sigma}p_{_V\mu}-p_{_V\sigma}\varepsilon_{_V\mu})
\varepsilon_{_{V^{'}\nu}}]\sum_{q=u,d,s}(D^{\nu}_2(x,m_q))
({1\over 2}p_{_V}-q)^\sigma m_Q\nonumber \\
&&+if_{_{V^{'}}}m_{_{V^{'}}}if_{_V}M_{_V}[2\varepsilon_{_V\mu}\varepsilon_{_{V^{'}\nu}}]
\sum_{q=u,d,s}(D^{\nu}_2(x,m_q))m_Q^2,
 \}\nonumber
\end{eqnarray}
and
\begin{eqnarray}
&&\langle V^{'}\gamma|M_{B}|V\rangle \nonumber \\
&&=i\pi^2eQ_Qg_s^4T^aT^b\otimes T^bT^a{1 \over (2\pi)^4}{1\over
({1\over 2}p_{_V}-q)^2-m_Q^2}{1\over
N_c^2}\varepsilon^{*\mu}(\gamma)\int_0^1dx\phi_P(\mu,x)\nonumber
\\
&&\{-if_{_{V^{'}}}m_{_{V^{'}}}if_{_V}M_{_V}[2g_{\nu\sigma}\varepsilon_{_V\mu}\varepsilon_{_{V^{'}\rho}}
+2g_{\sigma\mu}\varepsilon_{_V\nu}\varepsilon_{_{V^{'}\rho}}-
2g_{\mu\nu}\varepsilon_{_V\sigma}\varepsilon_{_{V^{'}\rho}}+
2g_{\rho\nu}\varepsilon_{_V\mu}\varepsilon_{_{V^{'}\sigma}}\nonumber
\\
&&+
2g_{\rho\nu}g_{\sigma\mu}\varepsilon_{_V\beta}\varepsilon_{_{V^{'}}}^\beta-
2g_{\rho\nu}\varepsilon_{_V\sigma}\varepsilon_{_{V^{'}\mu}}]\sum_{q=u,d,s}(D^{\rho\nu}_1(x,m_q))
({1\over 2}p_{_V}-q)^\sigma \nonumber \\
&&+if_{_{V^{'}}}m_{_{V^{'}}}if_{_V}i[
2i(\varepsilon_{_V\nu}p_{_V\mu}-p_{_V\nu}\varepsilon_{_V\mu})\varepsilon_{_{V^{'}\rho}}
+2ig_{\rho\nu}(\varepsilon_{_V\beta}p_{_V\mu}-p_{_V\beta}\varepsilon_{_V\mu})\varepsilon_{_{V^{'}}}^\beta
]\sum_{q=u,d,s}(D^{\rho\nu}_1(x,m_q))m_Q\nonumber \\
&&+if_{_{V^{'}}}m_{_{V^{'}}}if_{_V}i[2i(\varepsilon_{_V\sigma}p_{_V\mu}-p_{_V\sigma}\varepsilon_{_V\mu})
\varepsilon_{_{V^{'}\nu}}]\sum_{q=u,d,s}(D^{\nu}_2(x,m_q))
({1\over 2}p_{_V}-q)^\sigma m_Q\nonumber \\
&&-if_{_{V^{'}}}m_{_{V^{'}}}if_{_V}M_{_V}[2\varepsilon_{_V\mu}\varepsilon_{_{V^{'}\nu}}]
\sum_{q=u,d,s}(D^{\nu}_2(x,m_q))m_Q^2.
 \}\nonumber
\end{eqnarray}
\noindent{ Appendix B}\\
The two four-point loop functions in our text are:
\begin{eqnarray}
&& D^{\rho\nu}_1(x,m_q) = {1\over i\pi^2}\int d^4k
{(p_1+k)^\rho(p_2+k)^\nu\over
k^2[(p_1+k)^2-m_Q^2](k+p_2)^2[(k+p_3)^2-m_q^2]}, \nonumber \\
&& D^{\nu}_2(x,m_q) = {1\over i\pi^2}\int d^4k {(p_3+k)^\nu\over
k^2[(p_1+k)^2-m_Q^2](k+p_2)^2[(k+p_3)^2-m_q^2]},
\end{eqnarray}
with
\begin{eqnarray}
&&p_1 = {1\over 2}p_{_V}, p_2 = p_{_V}-q, p_3 = p_{_P} x.
\end{eqnarray}
And the three five-point loop functions are:
\begin{eqnarray}
&&E^{\nu\rho\sigma}_1(x,m_q) = {1\over i\pi^2}\int d^4k{k^\nu
k^\rho p_3^\sigma+k^\nu p_1^\rho
(k+p_3)^\sigma+p_4^\nu(k+p_1)^\rho(k+p_3)^\sigma\over
k^2[(k+p_3)^2-m_q^2](k+p_2)^2[(k+p_4)^2-m_Q^2][(k+p_1)^2-m_Q^2]},
\nonumber
\\
&&E^{\nu\sigma}_2(x,m_q) = {1\over i\pi^2}\int d^4k{(k+p_4)^\nu
(k+p_3)^\sigma\over
k^2[(k+p_3)^2-m_q^2](k+p_2)^2[(k+p_4)^2-m_Q^2][(k+p_1)^2-m_Q^2]},
\nonumber
\\
&&E^{\rho\sigma}_3(x,m_q) = {1\over i\pi^2}\int d^4k{(k+p_1)^\rho
(k+p_3)^\sigma\over
k^2[(k+p_3)^2-m_q^2](k+p_2)^2[(k+p_4)^2-m_Q^2][(k+p_1)^2-m_Q^2]},
\end{eqnarray}
with
\begin{eqnarray}
&&p_1 = {1\over 2}p_{_V}, p_2 = p_{_V}-q, p_3 = p_{_P} x, p_4 =
{1\over 2}p_{_V}-q,
\end{eqnarray}
which can be decomposed into somes of four-point loop functions
according to \cite{loop}.
\begin{eqnarray}
&&E^{\nu\rho\sigma}_1(x,m_q) = (\sum_{i=0}^{4}{det(Y_i)\over
det(Y)}D^{(fin)\nu\rho}(i)+\sum_{i,j=1}^{4}(-1)^{i+j}{det(\hat{Z}_{ij}^{(4)})\over
det(Y)}2p_{j\alpha}{\mathcal{D}}^{\alpha\nu\rho}(i))p_3^\sigma\nonumber \\
&&+(\sum_{i=0}^{4}{det(Y_i)\over
det(Y)}D^{(fin)\nu\sigma}(i)+\sum_{i,j=1}^{4}(-1)^{i+j}{det(\hat{Z}_{ij}^{(4)})\over
det(Y)}2p_{j\alpha} {\mathcal{D}}^{\alpha
\nu\sigma}(i))p_1^\rho\nonumber \\
&&+(\sum_{i=0}^{4}{det(Y_i)\over
det(Y)}D^{(fin)\nu}(i)+\sum_{i,j=1}^{4}(-1)^{i+j}{det(\hat{Z}_{ij}^{(4)})\over
det(Y)}2p_{j\alpha} {\mathcal{D}}^{\alpha \nu}(i))p_1^\rho
p_3^\sigma\nonumber \\
&&+(\sum_{i=0}^{4}{det(Y_i)\over
det(Y)}D^{(fin)\rho\sigma}(i)+\sum_{i,j=1}^{4}(-1)^{i+j}{det(\hat{Z}_{ij}^{(4)})\over
det(Y)}2p_{j\alpha}{\mathcal{D}}^{\alpha\rho\sigma}(i))p_4^\nu\nonumber \\
&&+(\sum_{i=0}^{4}{det(Y_i)\over
det(Y)}D^{(fin)\rho}(i)+\sum_{i,j=1}^{4}(-1)^{i+j}{det(\hat{Z}_{ij}^{(4)})\over
det(Y)}2p_{j\alpha}{\mathcal{D}}^{\alpha\rho}(i))p_4^\nu p_3^\sigma\nonumber \\
&&+(\sum_{i=0}^{4}{det(Y_i)\over
det(Y)}D^{(fin)\sigma}(i)+\sum_{i,j=1}^{4}(-1)^{i+j}{det(\hat{Z}_{ij}^{(4)})\over
det(Y)}2p_{j\alpha}{\mathcal{D}}^{\alpha\sigma}(i))p_4^\nu p_1^\rho\nonumber \\
&&+(\sum_{i=0}^{4}{det(Y_i)\over
det(Y)}D^{(fin)}_0(i)+\sum_{i,j=1}^{4}(-1)^{i+j}{det(\hat{Z}_{ij}^{(4)})\over
det(Y)}2p_{j\alpha}{\mathcal{D}}^{\alpha}(i))p_4^\nu p_1^\rho
p_3^\sigma, \nonumber \\
&&E^{\nu\sigma}_2(x,m_q) = (\sum_{i=0}^{4}{det(Y_i)\over
det(Y)}D^{(fin)\nu\sigma}(i)+\sum_{i,j=1}^{4}(-1)^{i+j}{det(\hat{Z}_{ij}^{(4)})\over
det(Y)}2p_{j\alpha}{\mathcal{D}}^{\alpha\nu\sigma}(i))\nonumber \\
&&+(\sum_{i=0}^{4}{det(Y_i)\over
det(Y)}D^{(fin)\nu}(i)+\sum_{i,j=1}^{4}(-1)^{i+j}{det(\hat{Z}_{ij}^{(4)})\over
det(Y)}2p_{j\alpha}{\mathcal{D}}^{\alpha\nu}(i))p_3^\sigma\nonumber
\\
&&+(\sum_{i=0}^{4}{det(Y_i)\over
det(Y)}D^{(fin)\sigma}(i)+\sum_{i,j=1}^{4}(-1)^{i+j}{det(\hat{Z}_{ij}^{(4)})\over
det(Y)}2p_{j\alpha}{\mathcal{D}}^{\alpha\sigma}(i))p_4^\nu
\nonumber \\
&&+(\sum_{i=0}^{4}{det(Y_i)\over
det(Y)}D^{(fin)}_0(i)+\sum_{i,j=1}^{4}(-1)^{i+j}{det(\hat{Z}_{ij}^{(4)})\over
det(Y)}2p_{j\alpha}{\mathcal{D}}^{\alpha}(i))p_4^\nu
p_3^\sigma, \nonumber \\
&&E^{\rho\sigma}_3(x,m_q) = (\sum_{i=0}^{4}{det(Y_i)\over
det(Y)}D^{(fin)\rho\sigma}(i)+\sum_{i,j=1}^{4}(-1)^{i+j}{det(\hat{Z}_{ij}^{(4)})\over
det(Y)}2p_{j\alpha}{\mathcal{D}}^{\alpha\rho\sigma}(i))\nonumber \\
&&+(\sum_{i=0}^{4}{det(Y_i)\over
det(Y)}D^{(fin)\rho}(i)+\sum_{i,j=1}^{4}(-1)^{i+j}{det(\hat{Z}_{ij}^{(4)})\over
det(Y)}2p_{j\alpha}{\mathcal{D}}^{\alpha\rho}(i))p_3^\sigma\nonumber
\\
&&+(\sum_{i=0}^{4}{det(Y_i)\over
det(Y)}D^{(fin)\sigma}(i)+\sum_{i,j=1}^{4}(-1)^{i+j}{det(\hat{Z}_{ij}^{(4)})\over
det(Y)}2p_{j\alpha}{\mathcal{D}}^{\alpha\sigma}(i))p_1^\rho
\nonumber \\
&&+(\sum_{i=0}^{4}{det(Y_i)\over
det(Y)}D^{(fin)}_0(i)+\sum_{i,j=1}^{4}(-1)^{i+j}{det(\hat{Z}_{ij}^{(4)})\over
det(Y)}2p_{j\alpha}{\mathcal{D}}^{\alpha}(i))p_1^\rho p_3^\sigma,
\end{eqnarray}
with :
\begin{eqnarray}
&&(Y)_{ij} = m_i^2+m_j^2-(p'_i-p'_j)^2,
\;\;\;\;\; i,j=0,\ldots,4, \nonumber \\
&&(\hat{Z}^{(4)})_{kl} =
2p'_k p'_l, \;\;\;\;\; k,l=1,\ldots,4, \nonumber \\
&&m_0 = 0, m_1 = m_q, m_2 = 0, m_3 = m_Q, m_4 = m_Q, \nonumber \\
&&p'_0 = 0, p'_1 = p_3, p'_2 = p_4, p'_3 = p_2, p'_4 = p_1.
\end{eqnarray}
And $D^{(fin)\nu\rho}(i)$'s denote the ultraviolet-finite
four-point functions that are obtained by removing the $i$th
propagator in the five-point functions. Let us take
$D^{(fin)\nu\rho}(1)$ as an example, and it is:
\begin{eqnarray}
&&D^{(fin)\nu\rho}(1) = {1\over i\pi^2}\int d^4k{k^\nu k^\rho
\over k^2(k+p_2)^2[(k+p_4)^2-m_Q^2][(k+p_1)^2-m_Q^2]}
\end{eqnarray}
where only four factors exist at the denominator.\\
$Y_i$ is obtained from the 5-dimensional Cayley matrix $Y$ by
replacing all entries in the $i$th column with 1 and the
3-dimensional matrices $\hat{Z}^{(4)}_{ij}$ result from the
4-dimensional Gram matrix $\hat{Z}^{(4)}$ by discarding the $i$th
row and $j$th column.

\pagebreak[4]
\begin{figure}[!htb]
\begin{center}
\caption{Radiative decays of a $^3S_1(Q\bar Q)$ bound-state into
$^1S_0(q \bar q)$ states in the lowest order.}
\begin{tabular}{cc}
{\includegraphics[width=12cm]{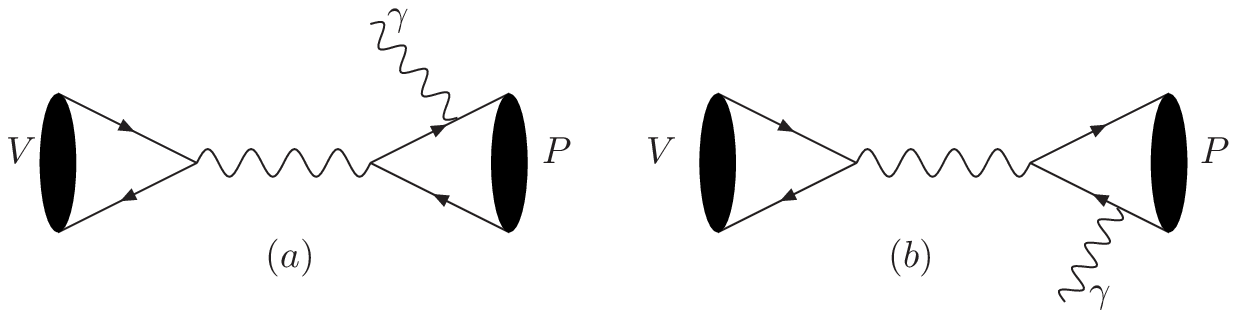}}
\end{tabular}
\end{center}
\label{fig1}
\end{figure}
\begin{figure}[!htb]
\begin{center}
\caption{Radiative decays of a $^3S_1(Q\bar Q)$ bound-state into
$^1S_0(q \bar q)$ states in the next order.}
\begin{tabular}{cc}
{\includegraphics[width=15cm]{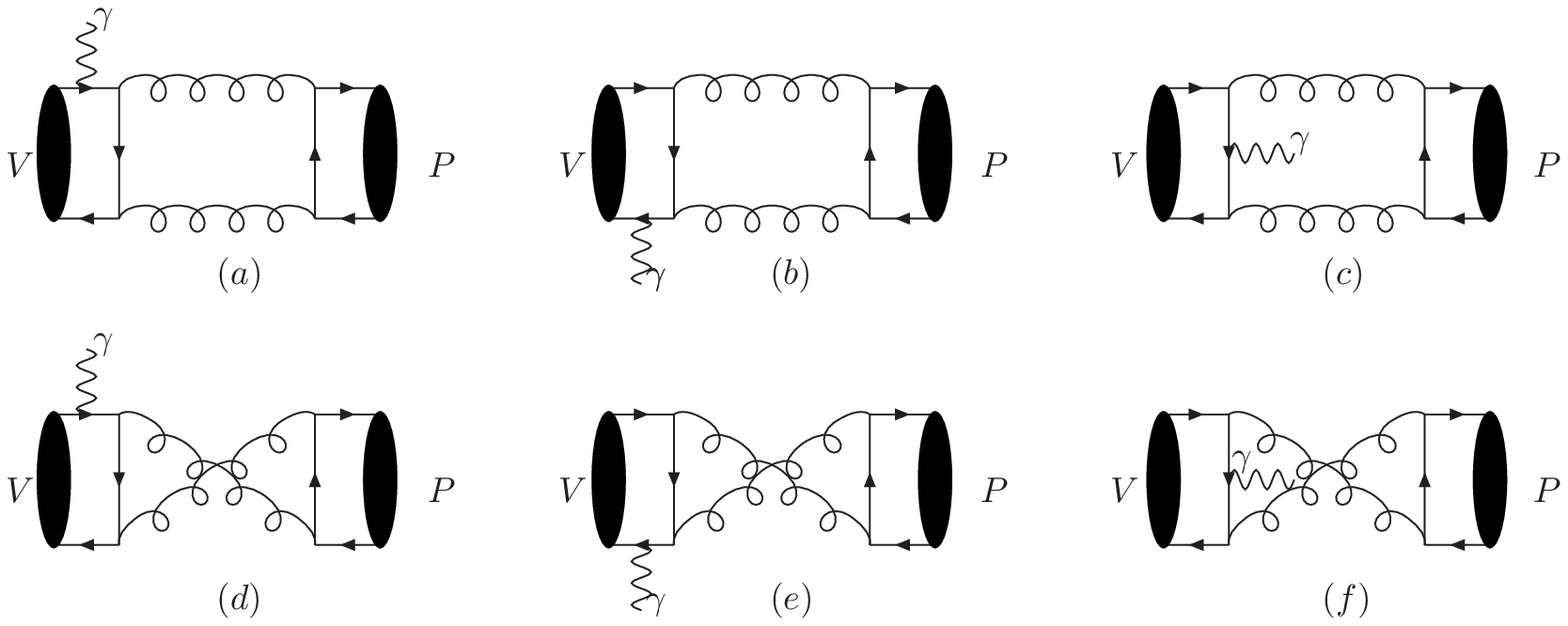}}
\end{tabular}
\end{center}
\label{fig1}
\end{figure}
\pagebreak[4]
\end{document}